\documentclass[manuscript]{aastex}

\slugcomment{Not to appear in Nonlearned J., 45.}

\shorttitle{The star cluster Trumpler~20}
\shortauthors{Carraro et al.}

\begin{document}


\title{Photometric characterization of the Galactic star
cluster Trumpler 20}


\author{Giovanni Carraro\altaffilmark{1}}
\affil{European Southern Observatory, Alonso de Cordova 3107, 
Casilla 19001, Santiago 19, Chile}
\email{gcarraro@eso.org}

\author{Edgardo Costa}
\affil{Departamento de Astronom\'ia, Universidad de Chile,
Casilla 36-D, Santiago, Chile}
\email{costa@das.uchile.cl}

\and

\author{Javier A. Ahumada}
\affil{Observatorio Astron\'omico, Universidad Nacional de C\'ordoba, Laprida 854, 
5000 C\'ordoba, Argentina}
\email{javier@oac.uncor.edu}


\altaffiltext{1}{On leave from: Dipartimento di Astronomia, 
Universit\`a di Padova, Italy}


\begin{abstract}
We present deep \emph{UBVI} photometry for Trumpler~20, a rich, intermediate-age
open cluster located at 
$l=301.47\degr$, $b=+2.22\degr$ ($\alpha=12^h~39^m~34^s$, 
$\delta=-60\degr~37^{\prime}~00^{\prime\prime}$, J2000.0) in the fourth Galactic quadrant.
In spite of its interesting properties, this cluster has received little attention,
probably because the line of sight to it crosses twice the
Carina spiral arm (and possibly also the Scutum-Crux arm), which causes a significant
contamination of its color-magnitude diagram (CMD) by field stars, therefore 
complicating seriously its interpretation.
In this paper we provide more robust estimates of the fundamental parameters of
Trumpler~20, and investigate the most prominent features of its CMD: a rich He-burning
star clump, and a vertical sequence of stars above the turnoff, which can be either
blue stragglers or field stars.
Our precise photometry, in combination with previous investigations, has allowed us to
derive updated values of the age and heliocentric distance of Trumpler~20, which we estimate
to be 1.4 $\pm$ 0.2~Gyr and 3.0 $\pm$ 0.3~kpc, respectively. As predicted by models, at this
age the clump has a tail towards fainter magnitudes and bluer colors, thus providing further
confirmation of the evolutionary status of stars in this particular phase. The derived heliocentric
distance places the cluster in the inter-arm region between the Carina and Scutum
arms, which naturally explains the presence of the vertical sequence of stars (which was
originally interpreted as the cluster itself) observed in the upper part of the CMD.
Most of these
stars would therefore belong to the general galactic field, while only a few of them
would be {\it bona fide} 
cluster blue stragglers.
Our data suggest that the cluster metallicity is solar, and that its reddening is \textit{E(B-V)}
$=$ 0.35 $\pm$ 0.04.
Finally, we believe we have solved a previously reported inconsistency between 
the spectroscopic temperatures and colors of giant stars in the cluster.
\end{abstract}



\section{Introduction}
Observational studies of Galactic open clusters have become a traditional benchmark
to test our comprehension of several aspects of stellar structure and evolution
(see Chiosi 2007, and references therein), and also of the formation and properties
of the Galactic disk (see Moitinho 2010, and references therein).\\
Being the clusters immersed in the Galactic general field, it is widely recognized
that, unless a detailed star by star membership analysis is available (which is
{\it not} the case for the vast majority of Galactic clusters, see Carraro et al.\ 2008),
the interpretation of their color-magnitude diagram (CMDs) is seriously complicated
by field stars located along
the line of sight to the cluster. Together with variable extinction, field star
contamination can produce sequences in the CMD which resemble typical cluster
sequences (especially in the case of very young clusters), leading to erroneous
interpretations. Unfortunately, the real nature of these field sequences can only
be clarified with a difficult {\it a posteriori} membership analysis (Villanova et al.\ 2005,
Moni Bidin et al.\ 2010).\\

This work is part of a series of papers aimed at improving the fundamental parameters
of poorly studied Galactic clusters (Seleznev et al. 2010; Carraro \& Costa 2007, 2009,
2010). Here we address the case of Trumpler~20, whose CMD
is obviously dominated by a significant field star population, which has been the
cause of past misinterpretations in regards to the cluster itself (Seleznev et al.\ 2010;
Platais et al.\ 2008 - hereafter Pla08; McSwain \& Gies 2005).\\ 

We present new, deep, \emph{UBVI} photometry, which has allowed us to put the fundamental
parameters of Trumpler~20 on a firmer base. We study the cluster's CMD in detail,
and investigate the nature of the conspicuous sequence of bright blue stars in
the upper CMD. This latter feature is common in clusters located at low Galactic latitudes,
and in this particular case its presence has led in the past to a misinterpretation of the 
cluster CMD (McSwain \& Gies 2005); here, moreover,
we address the question: are these stars blue stragglers (BS) that 
belong to the cluster or, more conservatively, are they simply field stars?
We also discuss the most prominent feature of the cluster CMD, namely its clump of He-burning
stars, and use it as a distance and age estimator. The clump is possibly the most
obvious indication that past classifications and basic parameters of Trumpler~20 
(particularly its age)
may be in error.\\

This paper is organized as follows. In Sect.~2 we summarize previous information
available for Trumpler~20. In Sect.~3 we present our observational material and
describe our reduction procedure. The cluster color-magnitude diagram is
 described in Sect.~4, while in
Sects.~5 to 7 we estimate its basic parameters. Sect.~8 is devoted to a discussion
on the cluster's clump, and Sect.~9 addresses the suspected BS population
of Trumpler~20. The global conclusions of the paper, together with suggestions
for future research directions, are given  in Section~10.

\section{The star cluster Trumpler~20 in perspective}
Trumpler~20 was first noticed by Trumpler (1930), who denoted it as An.\ 20.
He classified the cluster as a III~2r object, namely a
{\it detached cluster with no noticeable concentration}, with
{\it a medium range of brightness between the stars in the cluster}, and
{\it a rich cluster with over 100 stars}. Trumpler estimated a cluster
angular diameter of 10~arcmin, and a heliocentric distance of 2240~pc.
Decades later, Hogg (1965) also identified Trumpler~20 as an 8~arcmin cluster,
having 239 probable members down to $V \sim 17$~mag; and van den Bergh \& Hagen
(1975) described it as a real and rich cluster with an angular diameter of 
$\sim7$~arcmin, visible both in the blue and red plates of their homogeneous survey of
the southern sky.\\

More recently, Trumpler~20 was studied by McSwain \& Gies (2005), who
obtained Str\"omgren photometry down to $y = 17$~mag in the framework of a
search for Be stars in southern open clusters.  The sequence they recognized as
the main sequence of Trumpler~20 (see their Fig.~59) is however most probably composed
of field stars because, as recognized by Jean-Claude Mermilliod in the same year
(private communication to G.~Carraro), the cluster is much fainter. This prompted
an observational campaign which resulted in a much deeper \emph{VI} photometry acquired
in 2006, eventually published by Seleznev et al.\ (2010), which confirmed that
Trumpler~20 is actually an old cluster. The same misinterpretation was recognized
by Pla08, who secured \emph{BVI} photometry and redetermined the cluster
parameters by isochrone fitting, obtaining an age of $\sim1.3$~Gyr, for $E(B-V) = 0.46$,
and $(V-M_V)_{0}=12.15$. 
This age is consistent with the cluster's CMD, which
indeed shows a quite prominent clump, typical of
intermediate-age star clusters.
The Pla08 parameters are based on a spectroscopic
metallicity of [Fe/H]~$=-0.11$, derived from a single red giant star. The authors
mention, however,
that the value obtained for the reddening raises an inconsistency between
the photometric and spectroscopic temperatures. 
The distance obtained by Pla08 is 3.3 kpc, which
  puts the cluster much further away than Trumpler's
  early estimate.\\

\noindent
In Seleznev et al. (2010), we combined \textit{VI} optical photometry with 2MASS data
(Skrutskie et al.\ 2006), and focused our attention mainly on the structure of
Trumpler~20. Detailed star count analysis revealed that the cluster has a regular shape
and an angular diameter of 10~arcmin, confirming Trumpler's estimate based on a
visual inspection.  As shown in Fig.~7 of Seleznev et al.\ (2010), the radial density
profile is smooth, but the cluster shows a hole in its nominal center. Assuming
solar metallicity, we found a reddening consistent with the one derived by Pla08,
but a smaller distance of 2.9~kpc, for an age of 1.5~Gyr. Metallicity, together
with an insufficient color baseline, may explain these slightly different results.\\

\noindent
In an attempt to better characterize this interesting cluster, in 2009 we acquired new
deep \emph{UBVI} photometry. The description and interpretation of this photometric material
is the subject of this paper. We basically aimed at putting the cluster parameters on
a firmer base, and  tried to establish whether the blue sequence, 
erroneously indicated as the
main sequence of Trumpler~20 by  McSwain \& Gies (2005), is composed by field stars
or by cluster BS.\\

\noindent
As can be seen in Fig.~1, made from a 900~sec $I$-band exposure, Trumpler~20 is barely visible
in a very dense stellar field, which complicates the interpretation of its CMD (see below).
The field shown in Fig.~1 is 20~arcmin on a side; North is at the top, and the East to the left.

\section{Observations and Data Reduction}

\subsection{Observations}       

The region of interest (see Fig.~1) was observed with the Y4KCAM camera attached to the Cerro
Tololo Inter-American Observatory (CTIO) 1-m telescope, operated by the SMARTS 
consortium.\footnote{\tt http://http://www.astro.yale.edu/smarts} This camera is equipped with an STA
4064$\times$4064 CCD\footnote{\texttt{http://www.astronomy.ohio-state.edu/Y4KCam/ detector.html}}
with 15-$\mu$m pixels, yielding a scale of 0.289$^{\prime\prime}$/pixel
and a field-of-view (FOV) of $20^{\prime} \times 20^{\prime}$ at the
Cassegrain focus of the telescope. The CCD was operated without binning, at a nominal
gain of 1.44 e$^-$/ADU, implying a readout noise of 7~e$^-$ per quadrant (this detector is read
by means of four different amplifiers). 
\\
\noindent

In Table~1 we present the log of our \emph{UBVI} observations. All observations were carried out in
photometric, good-seeing, conditions. Our \emph{UBVI} instrumental photometric system was defined
by the use of a standard broad-band Kitt Peak \emph{UBVI$_{kc}$} set of filters.\footnote{\texttt{http://www.astronomy.ohio-state.edu/Y4KCam/ filters.html}}
To determine the transformation from our instrumental system to the standard Johnson-Kron-Cousins
system, and to correct for extinction, we observed 46 stars in Landolt's area SA~98 (Landolt 1992)
multiple times and with different
air-masses ranging from $\sim1.1$ to $\sim2.6$.  Field SA~98 is very advantageous, as it
includes a large number of well-observed standard stars, with a very good color coverage:
$-0.2 \leq (B-V)\leq 2.2$ and $-0.1 \leq (V-I) \leq 6.0$. Furthermore, it is completely covered by
the FOV of the Y4KCAM.

\subsection{Reductions}

Basic calibration of the CCD frames was done using the Yale/SMARTS y4k reduction script
based on the IRAF\footnote{IRAF is distributed
by the National Optical Astronomy Observatory, which is operated by the Association
of Universities for Research in Astronomy, Inc., under cooperative agreement with
the National Science Foundation.} package CCDRED. For this purpose, zero exposure
frames and twilight sky flats were taken every night.  Photometry was then performed
using the IRAF DAOPHOT and PHOTCAL packages. Instrumental magnitudes were extracted
following the point-spread function (PSF) method (Stetson 1987). A quadratic, spatially
variable, master PSF (PENNY function) was adopted. Aperture corrections were determined
making aperture photometry of a suitable number (typically 10 to 20) of bright, isolated,
stars in the field. These corrections were found to vary from 0.160 to 0.290 mag, depending
on the filter. The PSF photometry was finally aperture corrected, filter by filter.

\section{The photometry}

After removing problematic stars, and stars having only a few observations in Landolt's
(1992) catalog, our photometric solution for a grand total of 297 measurements per filter,
turned out to be:\\

\noindent
$ U = u + (3.080\pm0.010) + (0.45\pm0.01) \times X - (0.009\pm0.006) \times (U-B)$ \\
$ B = b + (2.103\pm0.012) + (0.27\pm0.01) \times X - (0.101\pm0.007) \times (B-V)$ \\
$ V = v + (1.760\pm0.007) + (0.15\pm0.01) \times X + (0.028\pm0.007) \times (B-V)$ \\
$ I = i + (2.751\pm0.011) + (0.08\pm0.01) \times X + (0.045\pm0.008) \times (V-I)$ \\

The final {\it r.m.s} of the fitting was 0.030, 0.015, 0.010, and 0.010 in $U$, $B$, $V$
and $I$, respectively.\

\noindent
Global photometric errors were estimated using the scheme developed by Patat \& Carraro
(2001, Appendix A1), which takes into account the errors resulting from the PSF fitting
procedure (i.e., from ALLSTAR), and the calibration errors (corresponding to the zero point,
color terms, and extinction errors). In Fig.~2 we present our global photometric errors
in $V$, $(B-V)$, $(U-B)$, and $(V-I)$ plotted as a function of $V$ magnitude. Quick
inspection shows that stars brighter than $V \approx 20$ mag have errors lower than
$\sim0.05$~mag in magnitude and lower than $\sim0.10$~mag in $(B-V)$ and $(V-I)$. Higher
errors are seen in $(U-B)$.\\

Our final optical photometric catalog consists of 13038 entries having \textit{UBVI} measurements
down to $V\sim 20$, and 43471 entries having \textit{VI} measures down to $V\sim 22$.

\subsection{Completeness}

Completeness corrections were determined by running artificial star experiments
on the data. Basically, we created several artificial images by adding artificial stars
to the original frames. These stars were added at random positions, and had the same
color and luminosity distribution of the true sample. To avoid generating overcrowding,
in each experiment we added up to 20\% of the original number of stars. Depending on
the frame, between 1000 and 5000 stars were added. In this way we have estimated that the
completeness level of our photometry is better than 50\% down to $V = 20.5$, and better than
90\% down to $V = 19.25$.

\subsection{Complementary infrared data and astrometry}

Our optical catalogue was cross-correlated with 2MASS, which resulted in a final catalog
including \textit{UBVI} and \textit{JHK$_{s}$} magnitudes. As a by-product, 
pixel (i.e., detector) coordinates
were converted to RA and DEC for J2000.0 equinox, thus providing 2MASS-based astrometry.\\

Using this \textit{VIJHK$_{s}$} catalog, Seleznev et al.\ (2010) performed a detailed star count
analysis, and derived the radial surface density profile and size of Trumpler~20. In 
this study the cluster's center was found to be at:
$\alpha=12^{h}~39^{m}~34^{s}$, $\delta=-60\degr~38^{\prime}~42^{\prime\prime}$;
and its diameter and core radius were determined to be $\sim 30$ arcmin and $\sim 5$ arcmin,
respectively.\\

In Sect. 5.1 we will use these values to estimate field star contamination in the CMDs.

\subsection{Comparison with previous photometry}

In Seleznev et al. (2010) we compared our older \textit{VI} photometry with that of Pla08, and found a 
good agreement both in $V$ and $(V-I)$. Here we present a comparison of our new \textit{BVI}
photometry, again with that published by Pla08, in $V$, $(B-V)$, and $(V-I)$. We note that Pla08 do not 
present $U$ photometry. Cross-correlating the two data sets we found 5373 stars in
common. The results of this comparison are plotted in Fig.~3.\\

As was found in Seleznev et al.\ (2010), the comparison is again good in both $V$ and
$(V-I)$.  Given that our photometry is much deeper, the significant scatter seen for $V$
fainter than $\sim 16.0$ is clearly due to the increasing errors at the faint tail of Pla08's
photometry. Here we find, however, an important difference in $(B-V)$. In general, this
could be due to a variety of reasons, but in this case we believe that the most probable
cause is the observing conditions under which the photometry of Pla08 was obtained.
These authors admit that they observed few standard stars -with a
quite narrow color range- at relatively high airmass. Together with $U$, the $B$ filter
is traditionally the most sensitive to observing conditions and the set of standard stars
used. The quite narrow color range can also explain the trend in the $V$ mag comparison,
which shows the presence of a shallow un-accounted color term.\\

As discussed later, this discrepancy could explain the difference we find in $E(B-V)$,
and the inconsistency between spectroscopic temperature and color discussed by Pla08.

\section{Color-magnitude diagrams}

In Fig.~4 we present the color-magnitude diagrams (CMDs) of Trumpler~20, 
based on all measured stars having photometric errors lower than
0.05 magnitudes, for three different color combinations: $V$ vs.\ $(U-B)$, 
$V$ vs.\ $(B-V)$ and $V$ vs.\ $(V-I)$.\\ 

\noindent
These CMDs are clearly dominated by dwarf stars (the conspicuous main sequence - MS) and 
giant stars from the thin disc (notice the sequence departing from the MS at $V\sim19$--20), 
located at different distances, and affected by different amounts of extinction. The FIRB
reddening in the line of sight (Schlegel et al. 1998) is $E(B-V) = 1.09$, which implies
$A_V \sim 3.0$. Given that the line of sight to Trumpler~20 crosses twice the Carina spiral
arm, and the Scutum-Crux arm (Russeil 2003), this reddening value (being an integration to infinity) is
probably much larger than the one at the distance of the cluster.\\

\noindent
A closer inspection of Fig.~4 shows that:

\begin{description}

\item $\bullet$ the CMD is dominated by a prominent, broad, MS, extending from the turnoff
point at $V\sim 16$ down to the limiting magnitude of our study;

\item $\bullet$ at V ~ 14.5 there is a conspicuous clump
  of He-burning stars, which extends significantly
  in magnitude;

\item $\bullet$ a sequence of bright blue stars is seen in the upper left part of the CMDs,
extending up to the saturation limit of our data;

\item $\bullet$ many field stars -dwarfs and giants- are spread across the CMD, which complicate
the precise definition of all the above features.
\end{description}

Overall, this CMD closely resembles that of NGC~7789, both in shape and
richness. We can say that Trumpler~20 looks like a twin of NGC~7789 (see Sect. 6).

\subsection{Clean color-magnitude diagrams}

We have selected cluster members on the basis of their distance from the cluster center.
For this, from the star count analysis of Seleznev et al.\ (2010) we adopted a 
cluster core radius of 5~arcmin.\\

{\it Clean} CMDs are shown in Fig.~5. Field star contamination is still present, but the
most important features of the CMDs stand out much better. Most of the stars above the TO have 
disappeared, which has allowed us to better define its position at:
$V$ = 16.0, $(B-V)$ = 0.75, $(V-I)$ = 0.85. While the MS in the $V$ vs.\ $(U-B)$ and
$V$ vs.\ $(B-V)$ CMDs are tight and separated from field stars and binaries, the 
$V$ vs.\ $(V-I)$ MS looks wide, and it appears impossible to separate the cluster's
MS from binaries and interlopers. Quite interestingly, the termination point of the MS
(the red hook) is still quite blurred, as if several distinct sub-populations were present.
We believe this is not the case, and will address this point below.

\section{Empirical determination of the fundamental parameters: comparison with NGC~7789}

Anticipating that a comparison of theoretical isochrones with Trumpler~20's CMD is very
complicated, due to a very important contamination by disk stars, 
we have applied an empirical method
to derive a first guess of the cluster fundamental parameters.\\

This exercise is illustrated in Fig.~6, where the left panel shows Trumpler~20's CMD, while
the middle panel shows that for NGC~7789, from Gim et al. (1998). We will concentrate on
these two panels for the moment.  Taking into account only their global shape, these two CMDs
look similar. They both have thick MSs, sequences of blue stars located along
the ideal continuation of the ZAMS, and prominent clumps. In none of them the MS TO is clear.
Disk giants are present in both CMD, although in the case of NGC~7789 they depart from the
MS at brighter magnitudes.  The differences seen in the precise location of the field stars
are the result of their different heliocentric distances and Galactic latitudes, and the different
run of interstellar extinction towards them.. In fact,
NGC~7789 is located 5.4 degrees below the formal Galactic plane, while Trumpler~20
is at 2.2 degrees above the plane.\\

\noindent
To make this comparison more quantitative and useful, in the rightmost panel of Fig.~6 we
have considered only stars located inside the core radius of Trumpler~20, and over-plotted
the ridge line for NGC~7789.  This latter has been shifted by $\Delta V = -0.2$ mag and
$\Delta (V-I) = -0.05$ mag. Given that the comparison is quite convincing, we can then assume -
as a working hypothesis- 
that Trumpler~20 has the same metal content as NGC~7789, namely solar (Gim et al. 1998). 
Under this assumption, it turns out that the apparent distance modulus of Trumpler~20 is
0.2 mag larger than that of NGC~7789, and that it is slightly more reddened. The reddening
of NGC~7789 is $E(V-I)=0.365$ (Gim et al. 1998), and its apparent distance modulus is
$(V-M_V)=12.2$ mag,
which therefore gives $E(V-I)\sim 0.40$ and $(V-M_V)\sim12.4$ mag for Trumpler~20. These
values imply a distance of $\sim 3$~kpc from the Sun for the latter. While the TO's are well
matched, the red clump of Trumpler~20 is slightly fainter and redder, which might imply a
lower age. We recall that the age of NGC~7789 is around 1.6~Gyr. We shall try to derive the
age of Trumpler~20 in Sect.~8.

\section{More on reddening and metallicity}

Additional insights on the reddening and metallicity of Trumpler~20 can be obtained from
the two-color diagram (TCD), shown in Fig.~7.  Again, we consider only stars within the
cluster core, and with photometric errors lower than 0.09~mag in both colors. The solid
line plotted
is an empirical Zero Age Main Sequence (ZAMS) from Schmidt-Kaler (1982), 
along which we indicate
a few relevant spectral types. The dashed sequence is this same ZAMS, but shifted by
$E(B-V) = 0.35$ along the reddening vector (arrow in the bottom left corner of the plot).
The fit is reasonable for this value of the reddening, further confirming the results of
the previous Section.\\

From the TCD we can estimate 
the cluster metallicity by means of the ultraviolet excess index: 
$\delta_{0.6} = \delta(U-B) {(B-V)_0=0.6}$ (see Sandage 1969; Karatas \& Schuster 2006;
Carraro et al. 2008).
In our TCD, spectral type F stars lie in the range $0.85 \leq (B-V) \leq 1.0$.
We therefore need to  look at $(B-V)\approx 0.95$ in this diagram, and identify
stars whose mean deviation from the ZAMS color is $\delta_{0.6}$.  At color
$(B-V) =0.95\pm0.05$ we have identified 17 stars that fulfill this condition.
Despite the scatter, this values implies [Fe/H] $\sim -0.05\pm0.13$, that is,
almost solar metal abundance.

\section{Fitting theoretical isochrones to the CMD}

In Fig.~8 we have over-plotted solar metallicity theoretical isochrones, from the
Padova suite of models (Girardi et al.\ 2000a), on our $V$ vs.\ $(B-V)$ CMD.
Lacking any solid estimate of the metal content of Trumpler~20 we have
conservatively adopted a solar metal content (we remind the reader that the value given
by Pla08 ([Fe/H] = -0.11, Sect. 2) was obtained from spectroscopy of only one red giant star). We note
additionally that the metal content of the {\it twin} cluster NGC~7789 (Gim et al. 1998) is almost solar.\\

Adjusting isochrones to a CMD is not an easy and straightforward task, especially in
cases like that of Trumpler~20, where contamination from field stars plays an important
role. In spite of this, as shown in the left panel of Fig.~8, a fit based on the set of 
parameters discussed previously: a reddening of 0.35 mag, a visual apparent distance modulus
of 13.7, and an age of 1.4~Gyr, matches the cluster MS very well all the way down to our
limiting magnitude. We estimate (by eye inspection) that the uncertainties in $E(B-V)$
and $(V-M_V)$, for this value of the age, are about 0.04 and 0.1, respectively.  The
reddening corrected distance modulus is therefore 12.6 mag; within the uncertainties close
to the value
derived from the comparison with NGC~7789.\\

The TO is reasonably accounted for, while the isochrone clump has the correct magnitude,
but a slightly redder color. We believe this is a problem of the models, which possibly
rely on poor transformation from the theoretical to the observation plane, and on an
imperfect calibration of the mixing length parameter (Carraro  \& Costa 2007, Palmieri
et al.\ 2002, Moitihno et al.\ 2006).\\

To better understand what is happening in the vicinity of the TO, in the right panel of
Fig.~8 we present a zoom of this region, where the same isochrone is plotted twice;
once for the same set of parameters as in the left panel, and a second version shifted by
0.7 mag to account for binary stars. Clearly, the broadening of the MS region is mostly
due to unresolved binaries, together with some unavoidable field star contamination
(see also the discussion in Sect.~9).

\section{The red clump}

As shown by Girardi et al.\ (2000b), the red clump in Galactic star cluster of this age
has a well defined shape, with an extension to lower magnitudes and bluer color. Given
that one of the most interesting features seen in the CMD of Trumpler~20 is its prominent
red clump, here we test if the quality of our photometry allows for a study of the detailed
morphology of the clump.\\

To this aim, we need a refined selection of the red clump members. We first tried to
perform a preliminary membership analysis using proper motion components from UCAC3
(Zacharias et al.\ 2010). This effort was not successful, and our conclusion is that this
catalog is not useful to study clusters at large distances from the Sun (3~kpc in the
case of Trumpler 20). We therefore used the standard procedure of selecting more
probable cluster members on the basis of their distance from the cluster center.\\

In Fig.~9 we show a zoom of the red clump region in the $V$ vs.\ $(B-V)$ CMD of
Trumpler~20, considering only stars within 5 arcmin from the cluster center. The red
clump of Trumpler~20 indeed shows a structure which closely resembles that of NGC~7789,
which we know has a similar age (Girardi et al 2000b, Fig~4a). In this figure we have
also plotted a model (evolutionary track) from Girardi \& Salaris (2001), adopting $E(B-V)=0.35$ and
$(V-M_V)=13.7$, as derived above. The fit is reasonable, and provides a further
confirmation of the age, reddening and distance we obtained in previous sections.\\

\noindent
As discussed by Girardi et al. (2000b), this morphology of the clump may be resulting 
either from star-to-star variations in the mass-loss rates during the RGB phase or by other
effects, such as stellar rotation or convective core overshooting, 
which can cause a significant spread in the core mass at He-ignition for stars of similar mass.
Apart from NGC~7789 and Trumpler~20, a similar morphology has been found in NGC~2204 and NGC~2660
(Girardi et al. 2000b).

\section{The sequence of blue bright stars: blue stragglers or field stars?}

The close similarity between the CMDs of Trumpler~20 and NGC~7789 also applies to the
population of bright blue stars. These stars can be either field stars
located between the cluster and the observer, or blue stragglers (Ahumada \&
Lapasset 2007). These latter should preferentially lie within the cluster area.
According to a recent study by Carraro et al.\ (2008), in the case of NGC~7789 it turns
out that most bright stars in this part of its CMD are interlopers, and only a minor
percentage are BSs. Here we investigate if the same scenario applies to Trumpler~20.\\

In Fig.~10 we present a $V$ vs.\ $(B-V)$ CMD of Trumpler~20, based only on stars
within the cluster's area, and indicate the region where, according to the
classic definition (see, e.g., Fig.~1 of Ahumada \& Lapasset 1995, 2007)
BSs should lie. In this diagram
we have over-plotted a ZAMS corresponding to its reddening and distance modulus (red solid
line), an isochrone corresponding to its age, reddening and distance (red dotted line),
and two straight blue segments indicating the probable location of BSs (see below for an
explanation of the red dashed line).  Inside this latter region we count 65 stars;
whether these objects are genuine BSs members of Trumpler~20 is hard to establish.\\

\noindent
Trumpler~20 lies very close to the northern border of the Coalsack dark nebula, in the 
northern edge of the Carina arm. In this direction, Russeil et al.\ (1998) found several
groups of young stars, three star clusters (NGC~4755, NGC~4463, and NGC~4439), and two
HII regions (RCW~69 and RCW~71), all at distances between 1.6 and 2.2~kpc (that is, closer
than Trumpler~20), which are consistent with the heliocentric distance and size of the
Carina spiral arm.\\
We note that the reddening in these directions to the Carina arm is about 0.35 mag.\\
We may therefore expect that most of the stars close to the green dashed ZAMS in Fig.~10
are stars located inside the arm. We note that this ZAMS has been displayed for the mean
distance and reddening of the Carina arm (2 kpc and 0.35 mag, respectively), at the
longitude of Trumpler~20. Interestingly, this line also crosses the TO region, implying
that stars from the Carina arm are significantly blurring the TO region.\\

We stress that what we are providing here is a mere qualitative description. Only a
detailed membership analysis will clarify the real percentage of BSs and field stars.

\section{Conclusions}

We have presented deep \textit{UBVI} and wide-field photometry for Trumpler~20, a rich open star cluster,
heavily contaminated by field stars, which lies inside the solar ring and in the inter-arm
region between Carina and Scutum-Crux. We have exploited our dataset aiming to improve
our knowledge of the cluster basic parameters. Having repeatedly stressed the crucial
role in the interpretation played by high contamination due to field stars, we conclude that
Trumpler~20 has an age of $1.4\pm0.2$~Gyr, making it a twin of the better-known open cluster 
NGC~7789.\\

\noindent
As anticipated in the Introduction, Galactic open clusters are ideal laboratories to test
theories of stellar evolution, and to probe Galactic structure.
Trumpler~20 appears to be quite a promising confirmation of this.\\

\noindent
On the stellar evolution side, we have shown that Trumpler~20 falls in the age range where
the clump of He burning stars exhibits a peculiar morphology, most 
possibly due to mass-loss variation during the RGB evolutionary phase. Other clusters of this age,
like NGC~2660, NGC~2204 and NGC~7789, are known to have a clump with the same morphology.\\

\noindent
On the Galactic structure side, we position Trumpler~20 in the inter-arm region
between the Carina and Scutum-Crux arms.
We remind the reader that not many clusters of this age are present in the inner
disk, possibly because of environmental effects,
which prevents survival of open clusters for a long time (Carraro et al. 2005).\\

\noindent
In this respect we believe that a proper spectroscopic study, to better assess membership
and metal content, would be really welcome for Trumpler~20.
From our photometric study, we can only suggest that its metallicity is probably solar.\\
Knowledge of its metal abundance would be of paramount importance to help constrain the slope and evolution
of the radial abundance gradient in the inner disk -where Trumpler~20 lies- which
has yet to be explored (Magrini et al. 2009, 2010).

\acknowledgments
EC acknowledges support by the the Chilean Centro de Astrof\'{\i}sica 
(FONDAP No. 15010003) and the Chilean Centro de Excelencia en Astrof\'{\i}sica y
Tecnolog\'{\i}as Afines (PFB 06).
JAA acknowledges ESO for granting a visitorship at ESO premises in Santiago,
where part of this work has been done.
In preparation of this paper, we made use of the NASA Astrophysics 
Data System and the ASTRO-PH e-print server. This research made
use of the WEBDA database operated at the Institute of Astronomy of 
the University of Vienna, Austria. This work also made extensive use of
the SIMBAD database, operated at the CDS, Strasbourg, France. 
This publication makes use of data products from the Two Micron All Sky
Survey, which is a joint project of the University of Massachusetts and
the Infrared Processing and Analysis Center/California Institute of 
Technology, funded by the National Aeronautics and Space Administration
and the National Science Foundation.

\clearpage

\begin{figure}
\plotone{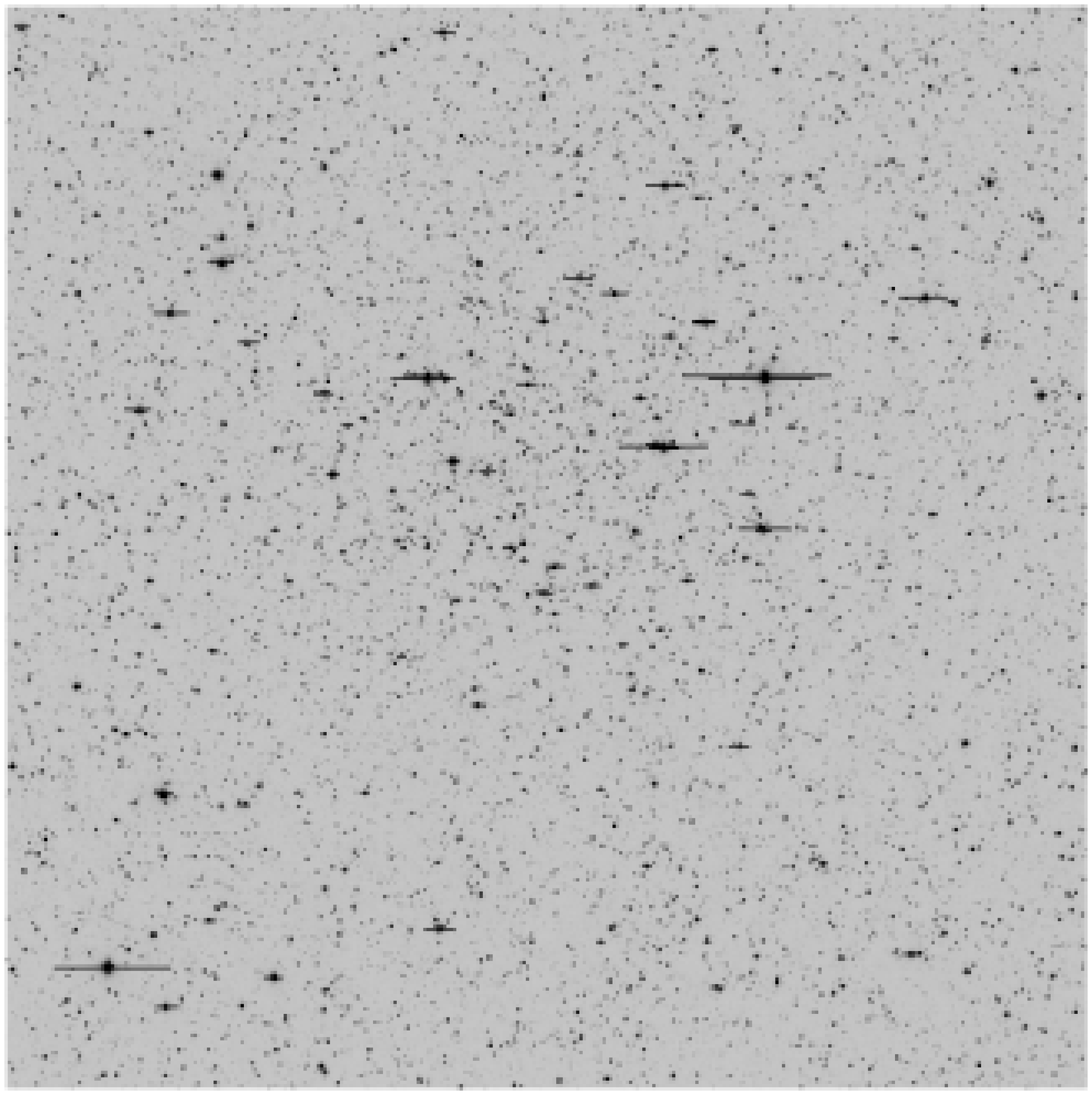}
\caption{$I$-band 900 sec image centered on Trumpler~20. The field is 20~arcmin on a side;
North is at the top, and East to the left.}
\end{figure}

\clearpage

\begin{figure}
   \plotone{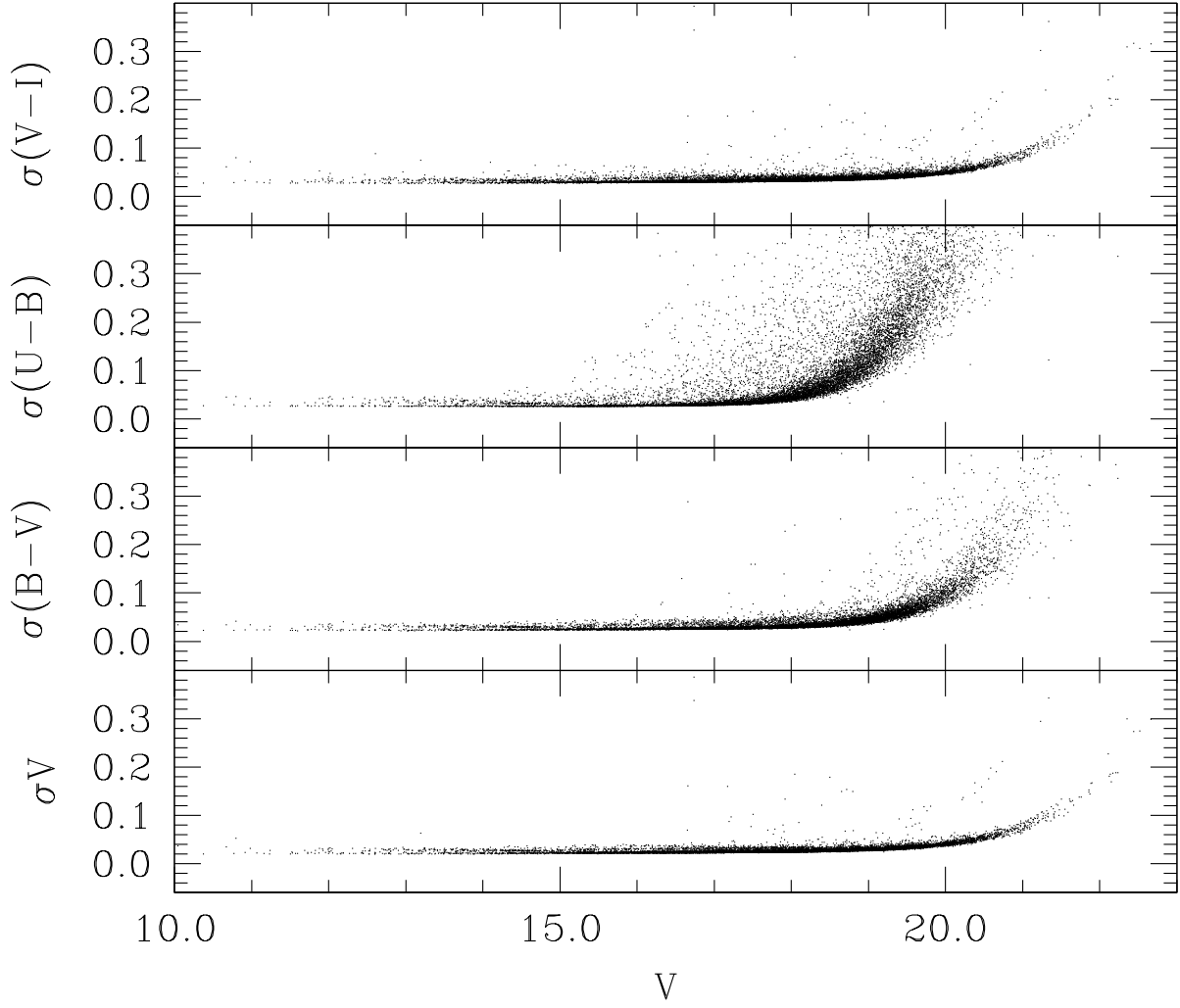}
   \caption{Photometric errors in $V$, $(B-V)$, $(U-B)$, and $(V-I)$ as a function of the $V$
            magnitude.}
   \end{figure}

\clearpage

\begin{figure}
   \plotone{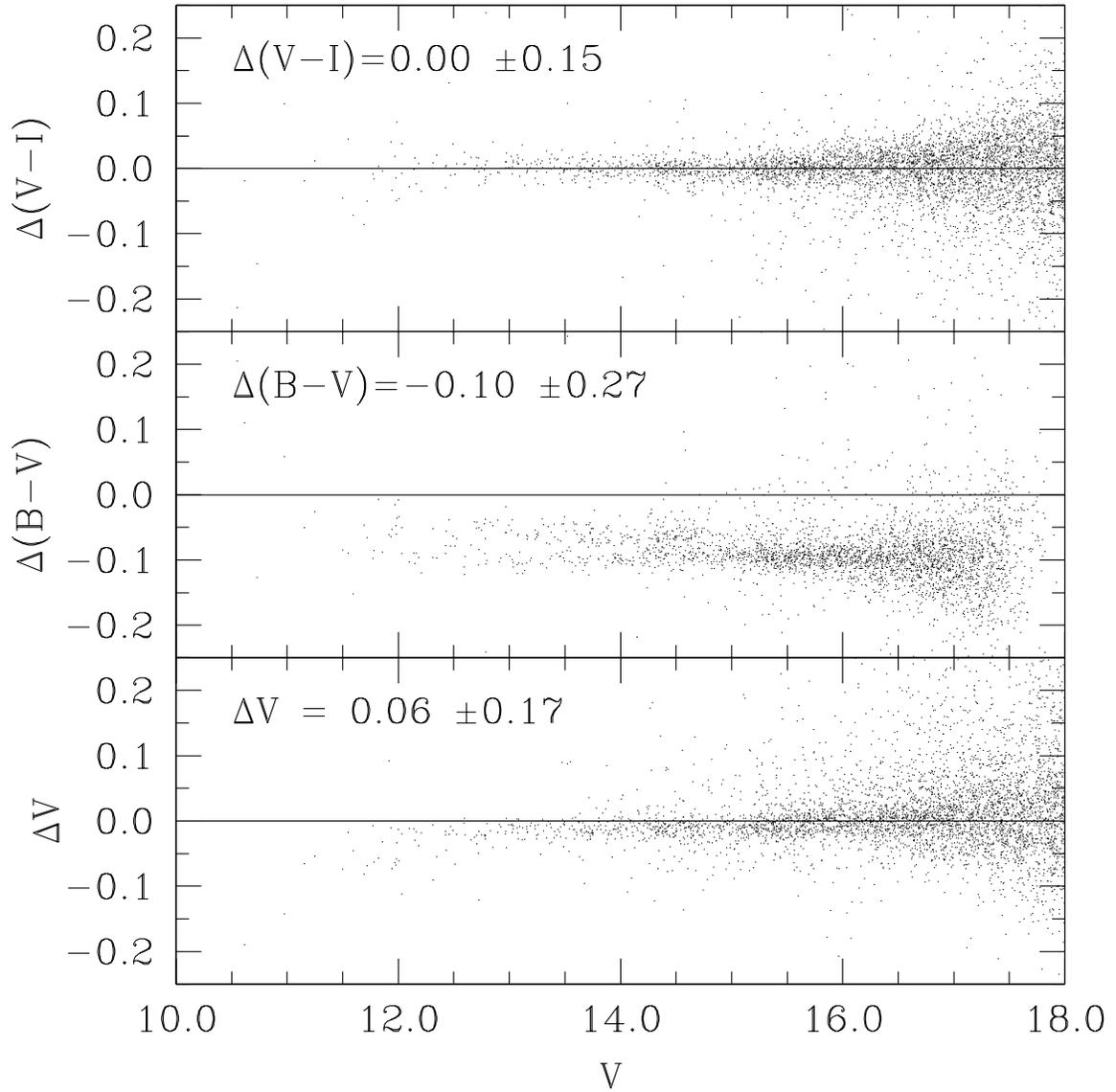}
   \caption{Comparison of our photometry with Pla08 for $V$, $(B-V)$ and $(V-I)$,
            as a function of $V$ magnitude. Comparison is in the sense our photometry
            minus Pla08.}
   \end{figure}

\clearpage

\begin{figure}
\plotone{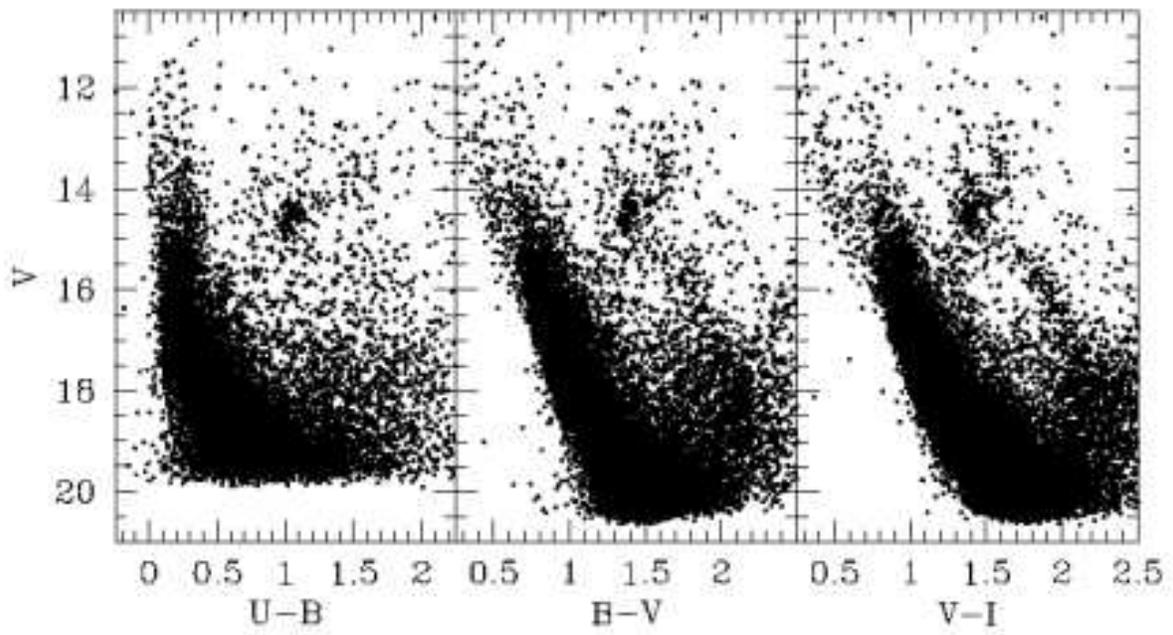}
\caption{Color-magnitude diagrams for three different color combinations, based on all
         measured stars having photometric errors lower than 0.05 magnitudes.}
\end{figure}

\clearpage

\begin{figure}
\plotone{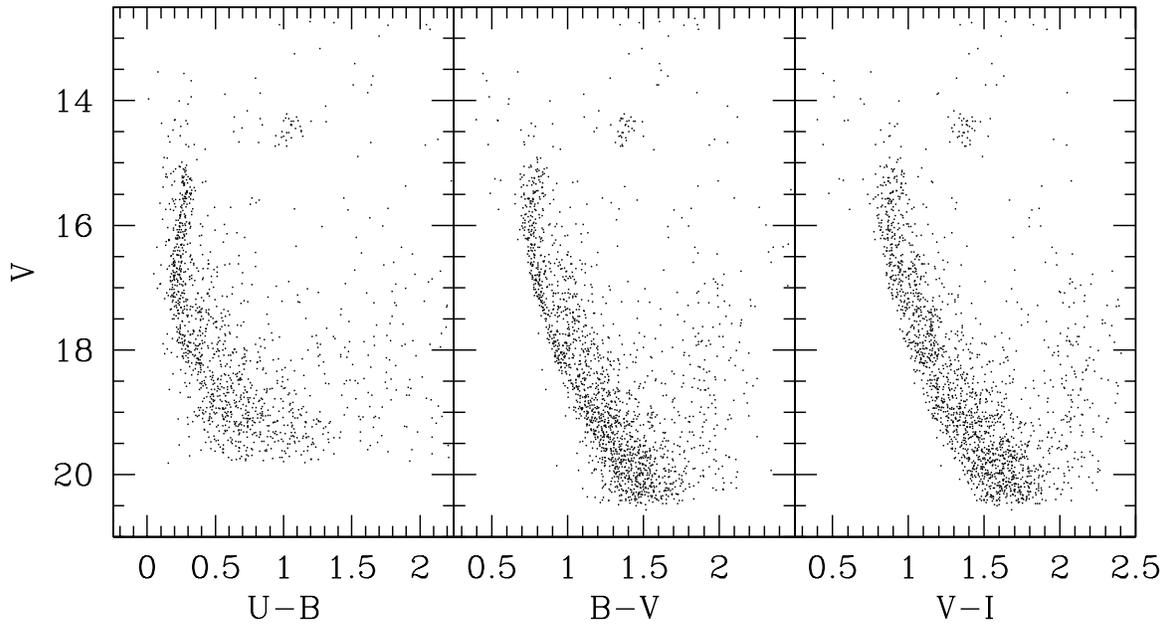}
\caption{Selection of cluster members on the basis of distance from the cluster center.
We have adopted a cluster core radius of 5 arcmin, from the star count analysis of 
Seleznev et al. (2010). The panels are the same as in Fig.~4.}
\end{figure}

\clearpage

\begin{figure}
\plotone{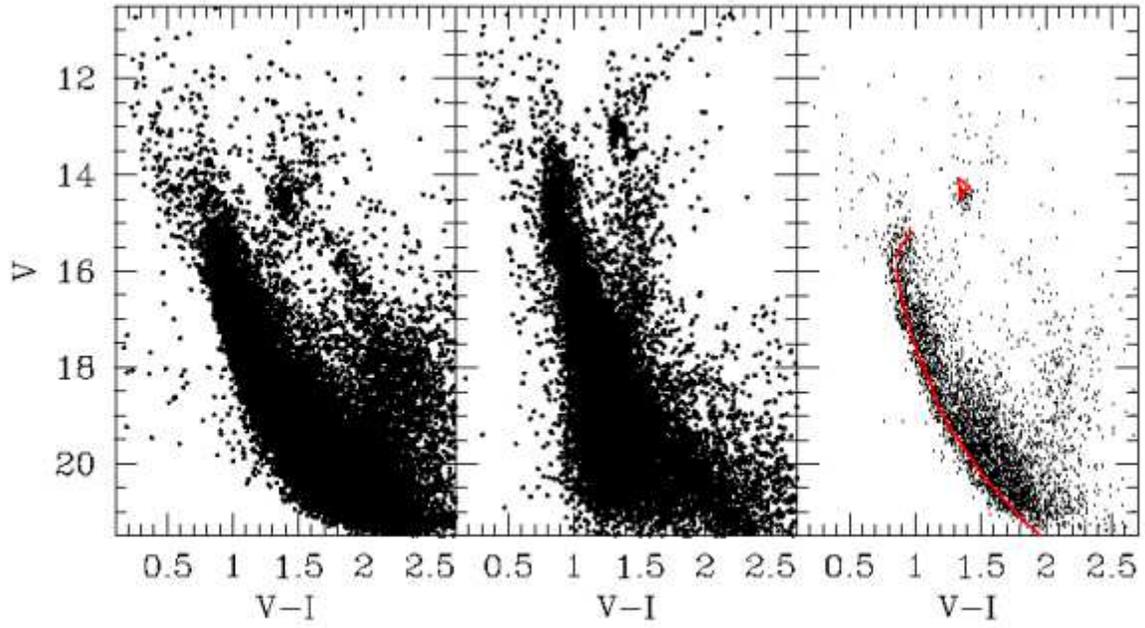}
\caption{Comparison with NGC~7789. {\bf Left panel}: Trumpler~20. {\bf Middle panel}: NGC~7789.
{\bf Right panel}: Trumpler~20 CMD for all stars within the core radius, with a superimposed
NGC~7789 ridge line.}
\end{figure}

\clearpage

\begin{figure}
\plotone{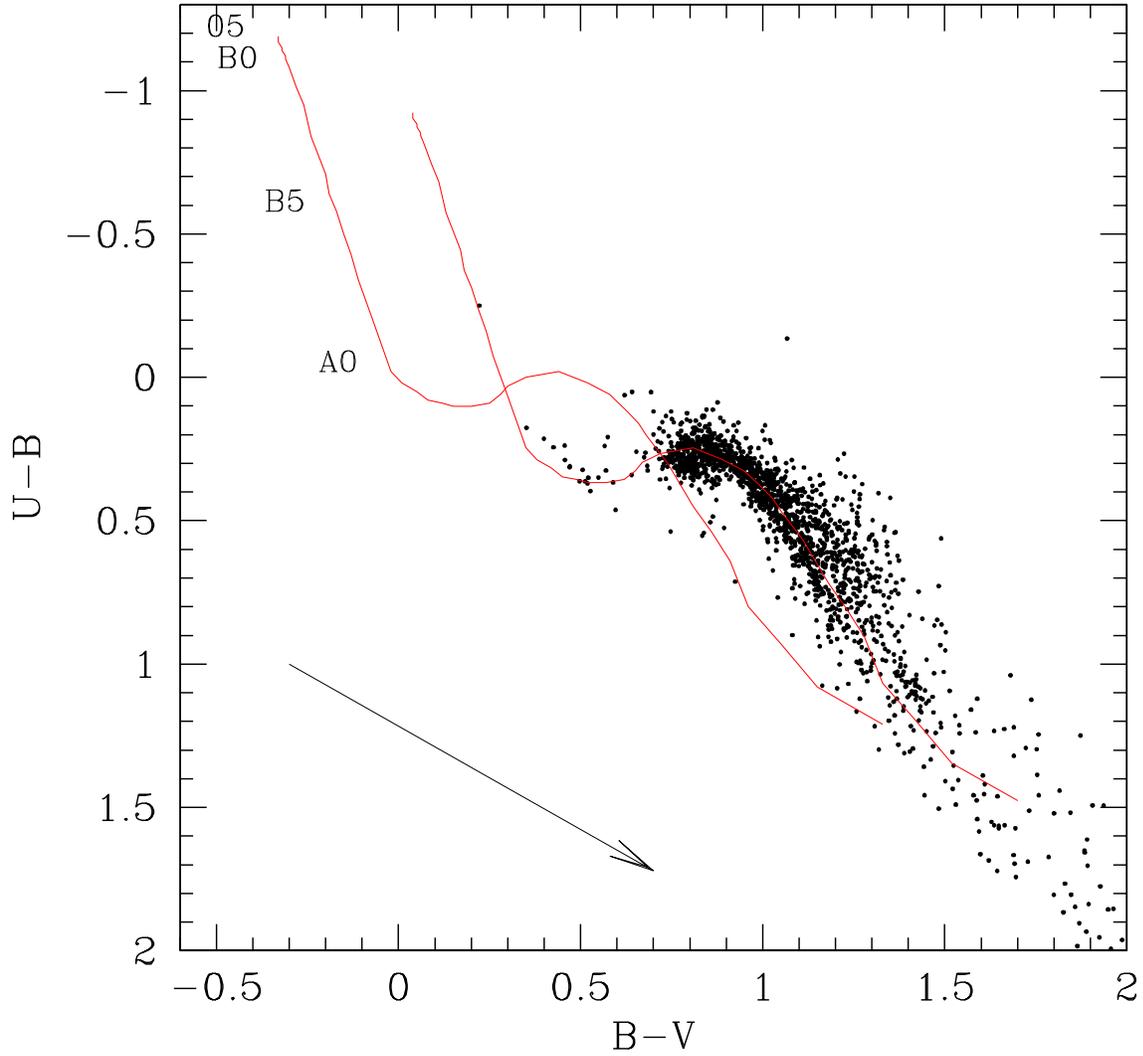}
\caption{Two color diagram for all stars within 5~arcmin from the center of Trumpler~20,
and having photometric errors lower than 0.09 mag in both colors. The solid and dashed lines
are empirical ZAMS for zero and 0.35 mag of $E(B-V)$. The normal reddening line is shown
in the lower left corner. For illustration purposes, a few spectral types are also
indicated.}
\end{figure}

\clearpage

\begin{figure}
\plotone{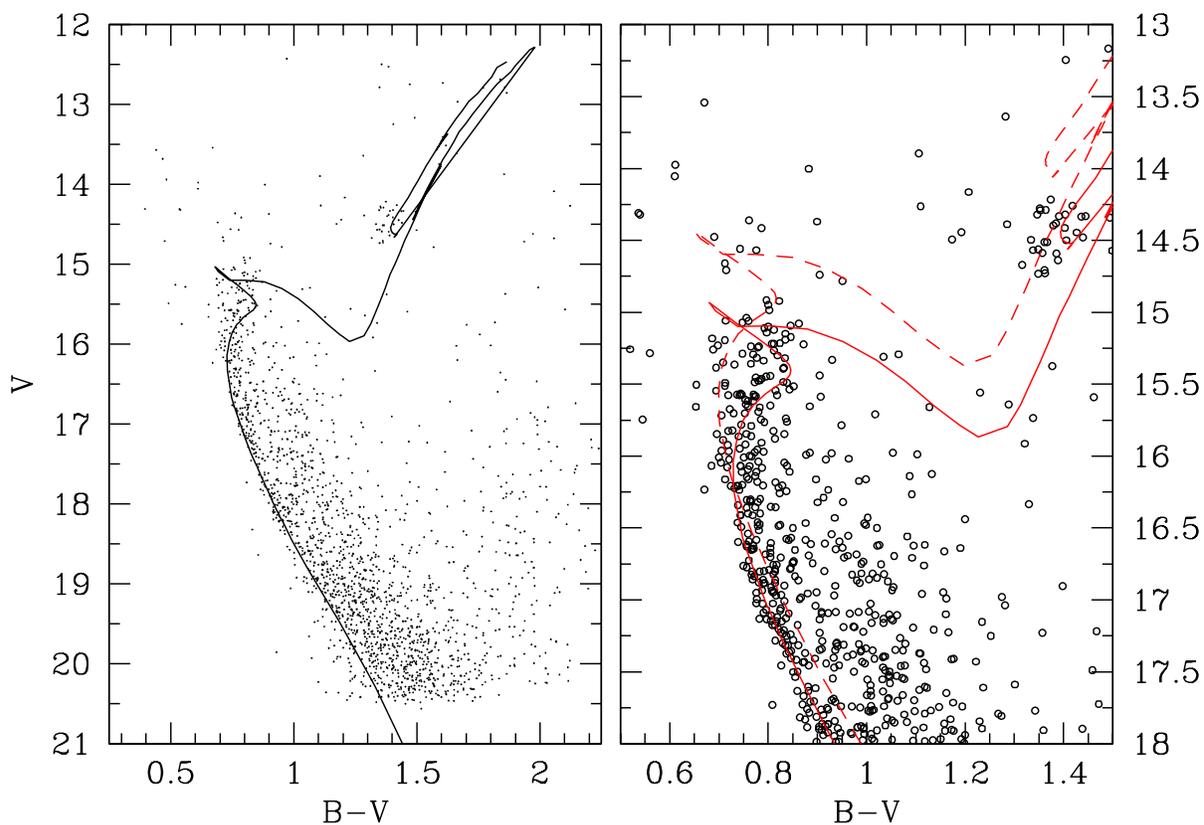}
\caption{{\bf Left panel}: Isochrone fitting to Trumpler~20 CMD, for the set of parameters
discussed in Sect~7, namely 0.35 mag, 13.7~mag and 1.4~Gyr for reddening, distance modulus
 and age,
respectively. {\bf Right panel}: a zoom of TO region. The solid line is the same isochrone as in
 the left
panel, while the dashed one is again the same isochrone, but shifted by 0.7 mag to account for
 unresolved
binary stars.}
\end{figure}

\clearpage

\begin{figure}
\plotone{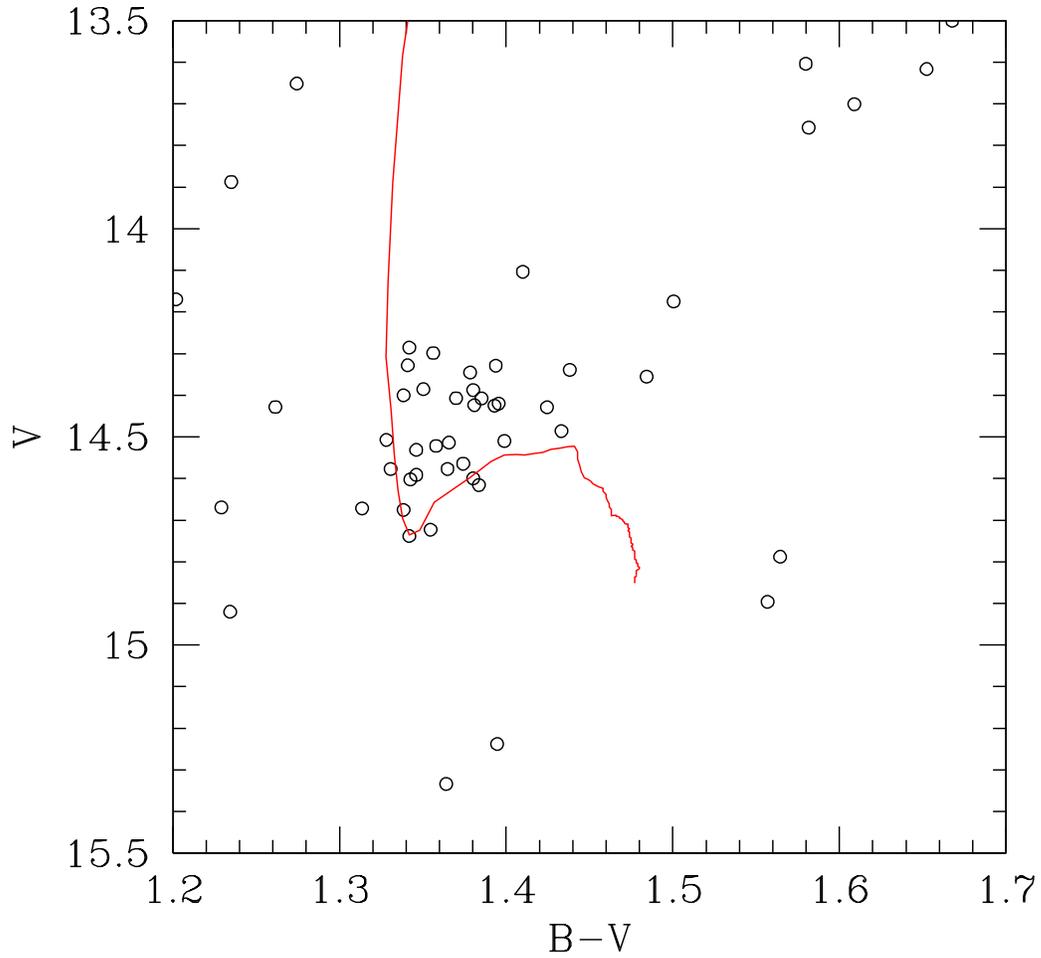}
\caption{Zoom of the red clump region in the $V$ vs.\ $(B-V)$ CMD of Trumpler~20.
To enhance the features, only stars within the cluster core radius ($\sim$5 arcmin) were
plotted. A model form Girardi \& Salaris (2001) has been over-plotted.}
\end{figure}

\clearpage

\begin{figure}
\plotone{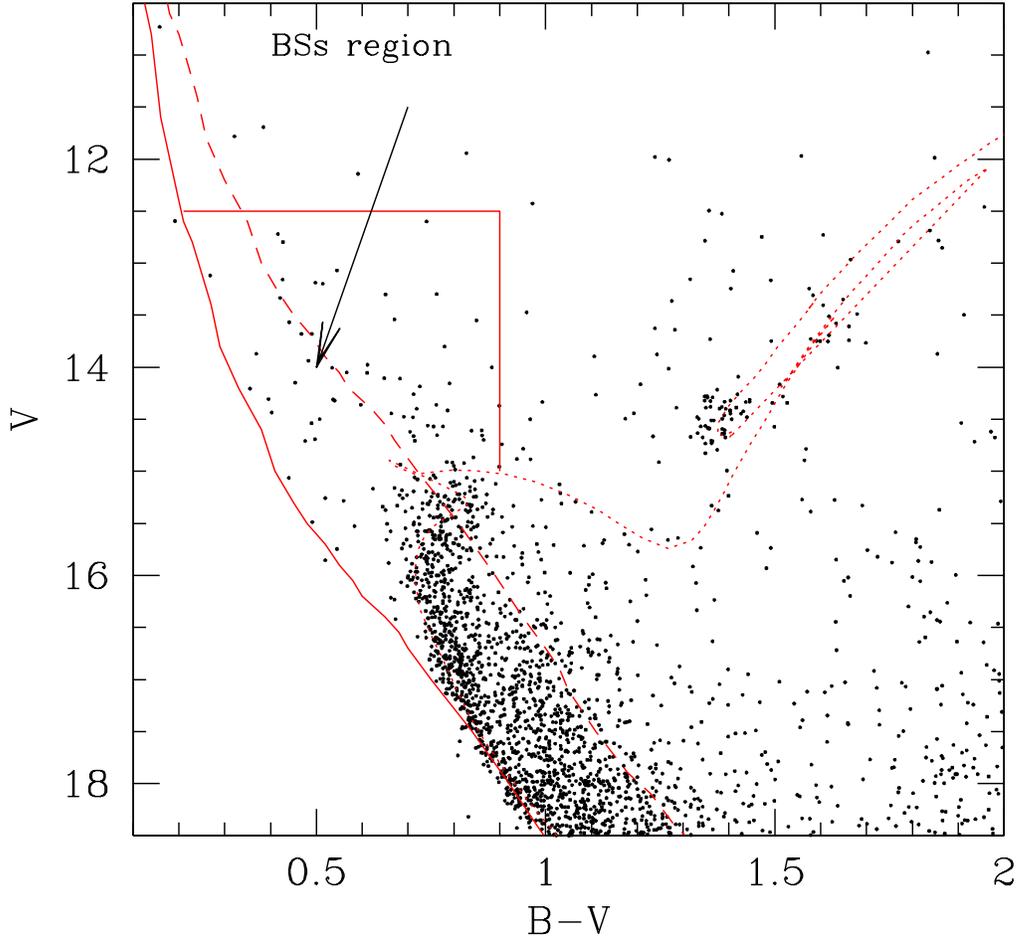}
\caption{$V$ {\it vs.} $(B-V)$ CMD of Trumpler~20 for stars within the core radius. The
solid red line is a ZAMS corresponding to the reddening and distance modulus of Trumpler~20,
while the red dotted line is an isochrone corresponding to its age, reddening and distance.
The region where BSs are expected to lie is delimited by two straight blue segments, and
indicated with an arrow. The green dashed line is a ZAMS displayed for the approximate
location and reddening of the Carina spiral arm, at the longitude of Trumpler~20.}
\end{figure}

\clearpage

\begin{table}
\tabcolsep 0.1truecm
\caption{$UBVI$ photometric observations.}
\begin{tabular}{lcccc}
\hline
\noalign{\smallskip}
Target& Date & Filter & Exposure (sec) & airmass\\
\noalign{\smallskip}
\hline
\noalign{\smallskip}
SA~98        & 2009 March 18   & \textit{U} & 2x20, 2x150, 2x400   &1.16$-$2.08\\
             &                 & \textit{B} & 2x20, 2x100, 2x200   &1.16$-$1.91\\
             &                 & \textit{V} & 2x10, 2x60, 2x120    &1.15$-$1.81\\
             &                 & \textit{I} & 2x10, 2x60, 2x120    &1.15$-$1.72\\
Trumpler~20  & 2009 March 18   & \textit{U} & 30, 200, 2000        &1.22$-$1.23\\
             &                 & \textit{B} & 20, 200, 1500        &1.28$-$1.29\\
             &                 & \textsl{V} & 10, 100, 900         &1.43$-$1.46\\
             &                 & \textsl{I} & 10, 100, 900         &1.36$-$1.38\\
PG~1047      & 2009 March 18   & \textit{U} & 2x30, 200            &1.49$-$1.52\\
             &                 & \textit{B} & 120                  &1.47\\
             &                 & \textit{V} & 20, 60               &1.40$-$1.42\\
             &                 & \textit{I} & 2x20, 60             &1.44$-$1.45\\
\noalign{\smallskip}
\hline
\end{tabular}
\end{table}
\end{document}